\documentclass[twocolumn,showpacs,amsmath,amssymb,aps,prl]{revtex4}

\usepackage{graphicx}
\usepackage{color}

\usepackage{amsmath}

\def\e{\epsilon}
\def\t0{\tau_0}

\newcommand{\be}{\begin{equation}}
\newcommand{\ee}{\end{equation}}
\newcommand{\bea}{\begin{eqnarray}}
\newcommand{\eea}{\end{eqnarray}}

\begin{document}

\title{Quantum Quench in the Transverse Field Ising Chain} 

\author{
Pasquale Calabrese$^1$, Fabian H. L. Essler$^2$, and Maurizio Fagotti$^{1}$
}
 \affiliation{$^1$Dipartimento di Fisica dell'Universit\`a di Pisa and
   INFN, Pisa, Italy\\
$^2$The Rudolf Peierls Centre for Theoretical Physics, Oxford
University, Oxford OX1 3NP, UK.}

\date{\today}

\begin{abstract}
We consider the time evolution of observables in the transverse field
Ising chain (TFIC) after a sudden quench of the magnetic field.
We provide exact analytical results for the asymptotic time and distance
dependence of one- and two-point correlation functions of the order
parameter. We employ two complementary approaches based on asymptotic
evaluations of determinants and form-factor sums.
We prove that the stationary value of the two-point correlation
function is not thermal, but can be described by a generalized
Gibbs ensemble (GGE). The approach to the stationary state can also be
understood in terms of a GGE. We present a conjecture on how these
results generalize to particular quenches in other integrable models.
\end{abstract}

\maketitle

Recent experiments on trapped cold atomic gases \cite{uc,kww-06} have raised
intriguing fundamental questions regarding the non-equilibrium dynamics of
correlated many-body quantum systems. These cold atom systems
are sufficiently weakly coupled to their environments as to allow the
observation of essentially unitary nonequilibrium time evolution on
long time scales. The quantum Newton's cradle experiments of 
Kinoshita et al \cite{kww-06} in particular have focussed the
attention on the roles played by dimensionality and conservation laws.
The observed absence of ``thermalization'' in quasi one
dimensional condensates was attributed to the experimental system
being approximately describable by a quantum integrable many-body theory.
This in turn initiated vigorous research on clarifying the role played
by quantum integrability in determining the stationary (late time)
behaviour of nonequilibrium evolution in correlated quantum systems
\cite{gg,cc-06,gge_various,fm-10,kla-06,bhc-10,gce-10}. 
The simplest way of driving a quantum system out of equilibrium is by
means of a {\it quantum quench}: a system is prepared in the ground
state of a given Hamiltonian $H(h_0)$, where $h_0$ is an experimentally
tuneable parameter such as a bulk magnetic field. At time $t=0$ the
parameter $h_0$ is changed suddenly to a different value $h$ and one
then considers the unitary time evolution of the system by means
of the new Hamiltonian $H(h)$. 
Central issues that have been investigated are whether the system
relaxes to a stationary state, and if it does, how to characterize its
physical properties at late times. It is widely believed (see
e.g. \cite{rev} for a comprehensive summary) that the behaviour of
local observables (such as one and two-point correlation functions)
can be described in terms of either an effective thermal (Gibbs)
distribution or a GGE\cite{gg}. It has been argued that the latter
arises for integrable models, while the former
obtains for generic systems\cite{gg,cc-06,gge_various,fm-10}.
However, several recent studies \cite{kla-06,bhc-10,gce-10} suggest
that the behaviour is more complicated and in particular depends on
the initial state. Moreover, open questions remain even with regards
to the very existence of stationary states. For example the order
parameter of certain mean-field models have recently been shown to
display persistent oscillations \cite{MF}.  

Two recent works have raised another crucial issue in the debate on
thermalization, namely the role played by the considered observables
\cite{bhc-10,can}. More precisely, it was pointed out that the
\emph{locality} of the observable with respect to the elementary
excitations is expected to affect the late time behaviour of an
observable after a quantum quench. In light of the available
experimental and theoretical results further clarification of
the role of integrability on the time evolution after a
quantum quench calls for {\it exact analytical} results on ``generic''
correlation functions, i.e. those corresponding to observables
non-local with respect to the elementary excitations. 

In the following we present such results for the particular case of
the transverse field Ising chain, which is a crucial paradigm for
quantum critical behaviour. While the model admits a representation in
terms of free fermions, the order parameter is non-local with respect
to the fermionic degrees of freedom, which renders it an ideal testing
ground for thermalization ideas. Although the model
has been widely analyzed in the context of quantum
quenches\cite{mc,ir-00,sps-04,fc-08,rsms-08}, the non-equilbrium
evolution of order parameter correlation functions are still not known
analytically. In this letter we present analytical results for the
{\it full asymptotic time and distance dependence} of one- and
two-point correlation functions of the order parameter in the
thermodynamic limit after a quantum quench within the ferromagnetic
phase. We also present partial results for quenches within the
paramagnetic phase and across the critical point. Our results are
obtained by two independent, novel methods. The first is based
on the determinant representation of correlation functions
characteristic of free-fermionic theories. The second is based on the
form-factor approach\cite{FF} and is applicable more generally to 
integrable quenches in interacting quantum field theories \cite{CEF}.
This method complements existing analytical/semi-numerical
methods used for studying quantum quenches in integrable systems
\cite{fm-10,INT}, but has the advantage of providing analytic answers
directly in the thermodynamic limit.

\begin{figure}[t]
\includegraphics[width=0.45\textwidth]{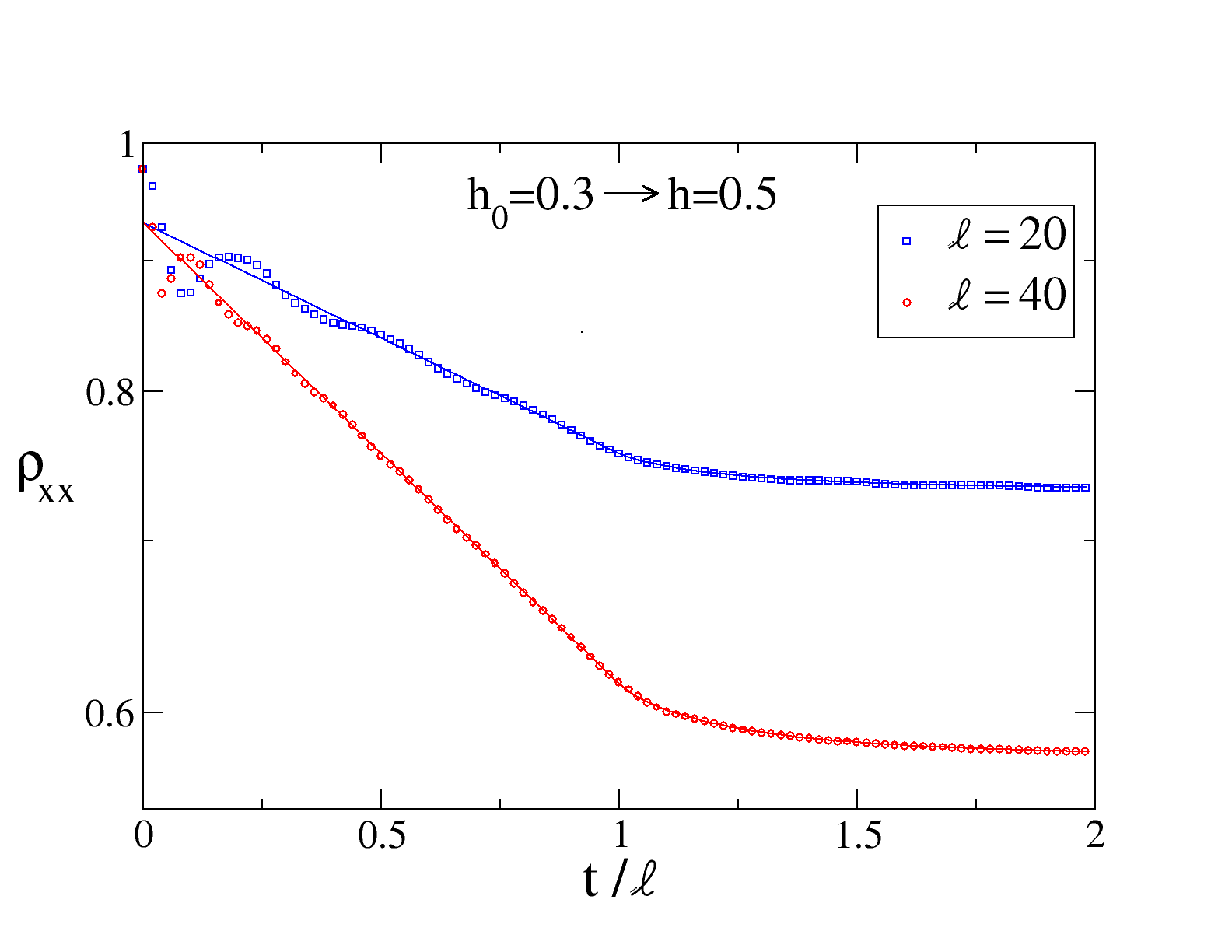}
\caption{$\rho^{xx}(\ell,t)$ for the quench $h_0=0.3\rightarrow h=0.5$
at fixed distance $\ell=20$ and $\ell=40$ against the  prediction in
\eqref{eq:prediction}. The overall amplitude of $\rho^{xx}$ has been
used as the same fit parameter in both cases.} 
\label{fig:1}
\end{figure}

\indent\emph{The model}. 
We consider the spin-\(\frac{1}{2}\) TFIC Hamiltonian
\be
H(h)=-\frac12 \sum_{l=-\infty}^\infty \left[
 \sigma_l^x\sigma_{l+1}^x+h \sigma_l^z \right]\,.
\ee
The Hamiltonian can be diagonalized by a combination of Jordan-Wigner
and Bogoliubov transformations (see e.g. \cite{mc}). The dispersion of
the elementary fermion excitations is $\e_h(k)=\sqrt{h^2 -2 h\cos k+1}$. 
The system is initially prepared in the ground state at a field
$h_0$. The field is then instantaneously changed from $h_0$ to $h$ and
unitary time evolution with Hamiltonian $H(h)$ ensues. We are
interested in the time evolution of the order parameter
$\rho^x(t)\equiv \langle{\sigma_l^x(t)}\rangle$ and its two-point
function $\rho^{xx}(\ell,t)\equiv
\langle{\sigma_l^x(t)\sigma_{l+\ell}^x(t)}\rangle$. Due to
translational invariance the 1-point function is position independent
and the 2-point function depends only on the distance $\ell$.
An important role is played by the difference $\Delta_k$ of the
Bogoliubov angles diagonalizing $H(h)$ and $H(h_0)$ respectively
\be
0<\cos \Delta_k=\frac{h h_0- (h+h_0) \cos k+1}{\e_h(k)
  \e_{h_0}(k)}\leq 1.
\ee
\indent{\it Quenches within the ordered phase ($h,h_0\leq 1$)}. 
We find that at late times the order parameter relaxes to zero
exponentially fast  
\begin{equation}
\label{eq:OP}
\rho^x(t) \propto\exp\left[t \int_{0}^\pi \frac{\mathrm d k}{\pi} \e'_h(k)
\ln\left(\cos\Delta_k\right) \right],
\end{equation}
where $\e'_h(k)=d\e_h(k)/dk$.
The two-point function of the order parameter exhibits exponential
decay both in time and distance ($\theta(x)$ denotes the Heaviside function)
%
\begin{multline}
\label{eq:prediction}
\rho^{xx}(\ell,t)\propto\exp\Bigg[\ell \int_0^\pi
\frac{\mathrm d k}{\pi}\ln\left(\cos\Delta_k\right)
\theta\big(2\e'_h(k)t-\ell\big)\Big]\\
\times\exp\Bigg[
2t \int_0^\pi\frac{\mathrm d k}{\pi} \e'_h(k)
\ln\left(\cos\Delta_k\right)\theta\big(\ell-2\e'_h(k)t\big)\Bigg].
\end{multline}
In the $\ell\to\infty$ limit the first factor is equal to unity
and $\rho^{xx}(\infty,t)=\big(\rho^{x}(t)\big)^2$, confirming
cluster decomposition in our non-equilibrium situation. 
Fig. \ref{fig:1} shows a comparison of our asymptotic result for
$\rho^{xx}(\ell,t)$ to numerical data, establishing the accuracy of
the former even for relatively short separations and times.
We note that (\ref{eq:prediction}) holds even for quenches to or from
the quantum critical point and agrees with the general form
put forward in \cite{cc-06} on the basis of semiclassical arguments.

\indent{\it The stationary state.} The result (\ref{eq:prediction})
allows us to make exact statements regarding thermalization in the model. 
The one-point function is trivially thermal, since it vanishes for
$t\to\infty$ as was already pointed out in \cite{cc-06,rsms-08}. 
On the other hand, in this limit the two-point function exhibits
exponential decay with a correlation length 
\be
\xi^{-1}= \int_{-\pi}^\pi \frac{dk}{2\pi}\xi^{-1}(k)=
-\int_{-\pi}^\pi 
\frac{\mathrm d k}{2\pi}\ln|\cos\Delta_k|\,.
\label{xiQ}
\ee 
This is reminiscent of the behaviour of the equilibrium two-point
function $\rho_{\rm eq}^{xx}(\ell)$ at temperature $T$, which
decays exponentially with correlation length (for $h<1$)\cite{mc}
\be
\xi^{-1}_T=
-\int_{-\pi}^\pi\frac{\mathrm d k}{2\pi} \ln\Bigl|\tanh\frac{ \epsilon_h(k)}{2T}\Bigr|\, .
\label{bm}
\ee
For $\rho^{xx}(\ell,\infty)$ to be thermal, 
$\xi$ would have to equal $\xi_{T_{\rm eff}}$, where the
effective  temperature $T_{\rm eff}$ is determined by the requirement
that the (average) energy in the initial state $\langle
\psi_0|H(h)|\psi_0\rangle$ is given by the thermal average
$\langle H(h)\rangle_{T_{\rm eff}}$. This leads to the following equation
fixing $T_{\rm eff}$
\be
\int_{-\pi}^\pi  \frac{dk}{2\pi} \e_h(k)\cos\Delta_k =
 \int_{-\pi}^\pi  \frac{dk}{2\pi}\e_h(k)\tanh\frac{\e_h(k)}{2T_{\rm eff}}\,.
\label{beff}
\ee
With $T_{\rm eff}$ given by (\ref{beff}) we find that
$\xi^{-1}_{T_{\rm eff}}\neq\xi^{-1}$ and $\rho^{xx}(\ell,\infty)$ is
therefore \emph{never} thermal. On the other hand, for small
quenches corresponding to small values of $T_{\rm eff}$, $\xi$ 
and $\xi_{T_{\rm eff}}$ can be seen to coincide to order $(h-h_0)^2$ 
\endnote{However the prefactors of the exponentials differ
in the two cases, which leads to deviations when $h$ is close to the 
critical point.}. Hence there exists a small quench regime, where the
thermal result provides a good approximation. This agrees with the
numerical findings of Ref. \cite {rsms-08}, which were reported to be
consistent with thermal behaviour at low effective temperatures.
In \cite{gg} it was proposed that the stationary state of integrable
models can be described in terms of a GGE, defined by maximizing the
entropy while keeping the energy as well as higher conservation
laws fixed. For the particular case of the TFIC this results in a 
{\it mode-dependent} effective temperature given by
\be
\cos\Delta_k = \tanh\frac{ \e(k)}{2T_{\rm eff}(k)}\,.
\label{eq:betah}
\ee
Inserting this relation into the expression (\ref{bm}) for $\xi_T$ in
Eq. (\ref{bm}) results in $\xi$ (\ref{xiQ}), which proves that the
stationary behaviour after a quench is indeed described by a GGE. 
To the best of our knowlege this demonstrates for the first time that
the GGE applies even to correlation functions of observables that are
nonlocal with respect to the elementary excitations in a lattice
model, thus significantly generalizing the results of
\cite{gg,cc-06,gge_various}. 
Our expression (\ref{xiQ}) for the correlation length in terms of
$T_{\rm eff}(k)$ (\ref{eq:betah}) explains nicely
observations made in \cite{rsms-08}, that numerical results on
$\rho^{xx}$ are described more accurately by a mode-dependent
correlation length than by the thermal result (\ref{bm}).
In \cite{fm-10} it was argued that the GGE quite generally describes
the stationary behaviour of one point functions for quenches in
integrable quantum field theories. Taking the appropriate scaling
limit of our result suggests the correctness of the assumptions made in
\cite{fm-10}. 

\indent\emph{Approach to the stationary state:}
Suprisingly the decay time in (\ref{eq:OP}) can be similarly
explained in terms of a GGE, even though it is not a property of the
stationary state. The one point function (\ref{eq:OP})
is characterized by exponential decay with rate
$\tau^{-1}=-\int_{0}^\pi \frac{\mathrm d k}{\pi} \e'_h(k)
\ln\left(\cos\Delta_k\right)$. This can be interpreted as the average
mode-dependent decay time $\tau^{-1}(k)=\e'_h(k)\xi^{-1}(k)$, obtained
by multiplying the mode-dependent inverse correlation length by the
velocity. The relaxational behaviour of the two-point function can be
understood following \cite{cc-06} by rewriting (\ref{eq:prediction})
as 
\be
\frac{\rho^{xx}(\ell,t)}
{\big(\rho^x(t)\big)^2}\sim \exp\left[\int_0^\pi\frac{dk}{\pi}
\Big[\frac{\ell}{\xi(k)}-\frac{2t}{\tau(k)}\Big]
\theta(2\e'_h(k)t-\ell)\right].
\label{2point}
\ee
The theta-function expresses the fact that a given mode can only
contribute to the relaxational behaviour if the distance $\ell$ lies
within its forward ``light cone'', while the form of the remaining
factor follows from the known stationary behaviour. Numerical studies
of the characteristic coherence time for the \emph{non equal-time} two
point function were found to be compatible with thermal behaviour
\cite{rsms-08}. It would be interesting to revisit this analysis in
light of our current findings.

{\it Quenches originating or ending in the disordered phase.}
Here the behaviour of correlation functions of $\sigma^x$ is more
involved \cite{rsms-08}. A complete summary of our results on the
dynamics in this case is beyond the scope of this note and will be
reported elsewhere \cite{CEF}. On the other hand, the correlation
length characterizing the stationary behaviour of
$\rho^{xx}(\ell,t=\infty)\sim\exp(-\ell/\xi)$  for an
arbitrary quench can be cast in the simple form
\begin{multline}
\xi^{-1}=\theta(h-1)\theta(h_0-1) \ln\left( {\rm min}[h_0,h_1]\right)\\
-\ln\left[x_++x_-+\theta\big((h-1)(h_0-1)\big)
\sqrt{4x_+x_-}\right],
\label{xigeneral}
\end{multline}
where $x_\pm=\frac{1}{4}[{\rm min}(h,h^{-1})\pm 1][{\rm
  min}(h_0,h_0^{-1})\pm 1]$
and $h_1=\frac{1+h h_0+\sqrt{(h^2-1)(h_0^2-1)}}{h+h_0}$. 
This agrees with (\ref{eq:prediction}) and the known results for
$h_0=0$ and $h_0=\infty$ \cite{sps-04}. Crucially, it can be proved by
a direct calculation in the framework of the Toeplitz determinant
approach summarized below that the result (\ref{xigeneral}) agrees
with the predictions of a GGE. We believe this calculation
generalizes straightforwardly to nonlocal correlators in other free
fermionic theories, which suggests that the GGE correctly predicts
infinite time behaviour for both local and nonlocal observables in
such theories. 

{\it Method I: Determinant approach}.
We focus on the two-point function $\rho^{xx}(\ell,t)$, which can be
written as the determinant of a $2\ell\times 2\ell$ block Toeplitz
matrix $T$ \cite{mc,rsms-08,CEF}. The matrix elements of $T$
depend explicitly on the time $t$. In the stationary state the $t$
dependence disappears \cite{sps-04} and the large-$\ell$ behaviour can
be obtained by application of the generalized Szego lemma\cite{CEF},
resulting in (\ref{xigeneral}). The dynamics in the limit
$t,\ell\to \infty$ at fixed ratio $t/\ell$ is much more difficult to
determine, as the elements of $T$ then depend on the matrix dimension
itself and Szego's lemma does not apply. To deal with this situation
we employ a method similar to \cite{fc-08}.
In order to calculate $\ln \det |T|= {\rm Tr}\,{\ln|T|}$, we consider
the moments of $T$, i.e.  ${\rm Tr}\, {T^{2n}}$ (we find that odd
moments are subleading). Calculating these moments gives
\begin{multline}
{\rm Tr}\,T^{2n}= \ell
\int_{-\pi}^\pi\frac{\mathrm d k}{2\pi} \big(\cos\Delta_k\big)^{2n}\\
+\int_{-\pi}^\pi\frac{\mathrm d k}{2\pi}
\varepsilon\left(\ell- 2|\e'_h(k)| t\right) 
\left[1-\big(\cos\Delta_k\big)^{2n}\right],
\end{multline}
where $\varepsilon(x)=x\theta(x)$.
The trace of any analytic function $f$ of $T$ can be formally expanded
in the moments. In our case we are interested
in $f(x)=\ln |x|$, which is analytic in the principal strip for any
$x\neq 0$.  As the symbol of the block Toeplitz matrix has winding
number zero about the origin and $\cos(\Delta_k)$ is always non zero we can
resum the expansion of the logarithm to obtain
(\ref{eq:prediction}). We have generalized these  results to the case
of the XY spin chain in a field \cite{CEF}.

{\it Method II: Form-factor approach}.
This approach applies more generally to quenches in integrable
(interacting) quantum field theories. We focus on the 1-point
function in the ordered phase. The ground state for $|h|<1$
spontaneously breaks the $\mathbb{Z}_2$ symmetry of the TFIC,
resulting in an initial (ground) state of the form
\be
|\Omega\rangle=\frac{1}{\sqrt{2}}\Big[|B\rangle_{\rm R}+
|B\rangle_{\rm NS}\Big],
\ee
where ${\rm R}$ and ${\rm NS}$ refer to the periodic/antiperiodic
sectors of the  free-fermion theory respectively and e.g.
$ |B\rangle_{\rm R}=\exp\left(
i\sum_{0<p\in{\rm R}}K(p) b^\dagger_p b ^\dagger_{-p}\right)|0\rangle_{\rm R}$,
where $K(p)=\tan\left[\Delta_p/2\right]$ and $b^\dagger_p$ is a
fermion creation operator with momentum $p$. The 1-point function
is
\be
\frac{\langle\Omega|\sigma^x_m(t)|\Omega\rangle}{\langle\Omega|\Omega\rangle}
=2\frac{{}_{\rm NS}\langle B|\sigma^x_m(t)|B\rangle_{\rm R}}{
{}_{\rm NS}\langle B|B\rangle_{\rm NS}+{}_{\rm R}\langle B|B\rangle_{\rm R}}.
\label{1point}
\ee
Expanding the ``boundary states'' $|B\rangle_{\rm R,NS}$ results in a
Lehmann representation for \eqref{1point}. Crucially, the matrix
elements (form factors) of $\sigma^x_m(t)$ between multifermion Hamiltonian
eigenstates are known exactly for the TFIC \cite{ising}. The main idea
for evaluating the Lehmann representation then follows the finite
temperature case \cite{AKT,EK09}. For a small quench the (total) density
$n_0$ of fermion excitations in the initial state constitutes a small
parameter. In this case one can use the $K(k)$-matrix as an expansion
parameter. One then observes that the form factors appearing in the 
Lehmann representation are singular when momenta in the in and out
states coincide. The leading (in the density $n_0$) contribution to
the one point fuction is obtained by summing all terms with the
strongest singularities at a given order in the expansion in powers of
$K$. This amounts to the exponentiation of infrared singularities. We
note that just as in the finite temperature case \cite{EK09} infinite
volume divergences encountered in evaluating the numerator of
(\ref{1point}) cancel against anologous divergencies in the denominator.
The result of these calculations is
\be
\frac{\langle\Omega|\sigma^x_m(t)|\Omega\rangle}{\langle\Omega|\Omega\rangle}
\propto
\exp\left[-t\int_0^\pi\frac{dk}{\pi}K^2(k)2\e'(k)\right]\,.
\label{final}
\ee
The decay rate agrees with the the leading term in the expansion of
(\ref{eq:OP}) in powers of $K^2(k)$. The correction to the $K^2(k)$
factor in (\ref{final}) are found to be ${\cal O}(K^6)$, again in
agreement with (\ref{eq:OP}). We note that (\ref{final}) provides an
excellent approximation to (\ref{eq:OP}) as long as $h,h_0$ are not
too close to the critical point. The two-point function can be
analyzed in an analogous manner \cite{CEF} and the results again agree
with the appropriate expansion of (\ref{eq:prediction}).

{\it Summary and Discussion}.
We have obtained exact analytic results for the long distance and time
asymptotic behaviour of one and two point functions of the order
parameter $\sigma^x$ in the TFIC atfer a quantum quench. We have shown
that the stationary expectation value of the two-point function is
not thermal, but can be described by GGE. We have further shown that the
approach to the stationary state for quenches within the ordered phase
can be understood in terms of a GGE as well. Our work further demonstrates
the importance of having analytic results at one's disposal when trying
to draw conclusions regarding the statistical description of
stationary state properties. 
Finally, we comment our newly developed method based on form
factors generalizes directly to \emph{integrable} quenches in
integrable quantum field theories. These are characterized by the
requirement that the initial state is compatible with factorizable
scattering. We conjecture that the stationary behaviour of both local
and nonlocal observables for integrable quenches in theories with
purely diagonal scattering, e.g. the sine-Gordon model at a
reflectionless point, can be described by an appropriate GGE.

\acknowledgments
We thank J. Cardy, D. Fioretto, G. Mussardo, D. Schuricht and A. Silva
for discussions. This work was supported by the EPSRC under grant
EP/D050952/1 (FHLE) by the ESF network INSTANS (PC and MF).


\end{document}